\documentclass[aps,
                    preprint
                    ]{revtex4}

\usepackage{epsfig}
\usepackage{wrapfig}
\usepackage{amsmath}
\usepackage{subfigure}

\def\e{\begin{equation}}
\def\f{\end{equation}}
\def\_#1{{\bf #1}}

\def\.{\cdot}

\begin{document}

\title{Transmission-line networks cloaking objects from electromagnetic fields}

\author{Pekka Alitalo$^1$, Olli Luukkonen$^1$, Liisi Jylh\"a$^2$, Jukka Venermo$^2$, Sergei Tretyakov$^1$}

\affiliation{$^1$Radio Laboratory / SMARAD Center of Excellence, TKK Helsinki University of Technology, P.O. Box 3000, FI-02015 TKK, Finland\\
$^2$Electromagnetics Laboratory, TKK Helsinki University of
Technology, Finland}

\begin{abstract}
We consider a novel method of cloaking objects from the
surrounding electromagnetic fields in the microwave region. The
method is based on transmission-line networks that simulate the
wave propagation in the medium surrounding the cloaked object. The
electromagnetic fields from the surrounding medium are coupled
into the transmission-line network that guides the waves through
the cloak thus leaving the cloaked object undetected. The cloaked
object can be an array or interconnected mesh of small inclusions
that fit inside the transmission-line network.

\end{abstract}


\maketitle{\center \large}

\section{Introduction}

Recently the subject of cloaking objects from electromagnetic
fields has aroused a lot of interest. Devices capable of cloaking
an object from the surrounding electromagnetic fields have been
recently suggested~\cite{Leonhardt,Pendry,Alu,Cai}. A realization
of an electromagnetic cloak, operating in the microwave regime has
been manufactured and measured~\cite{Schurig}. The operation of
the cloaks discussed in~\cite{Leonhardt,Pendry,Alu,Cai,Schurig} is
based on mapping of the electromagnetic fields in such a way that
the waves that come in contact with the cloak, are guided
\textit{around} the object which is placed inside the cylindrical
or spherical cloak. The main drawback of the designs presented
in~\cite{Leonhardt,Pendry,Alu,Cai,Schurig} is that these systems
are capable of cloaking from time-harmonic fields only. This is
because the waves that go around the cloaked object need to go as
fast as the wave propagating in free space on a straight line. No
signal can therefore be fully cloaked because the group velocity
in the (passive) cloak can never exceed the speed of light. Also
another approach to cloak design, based on hard surfaces, was
proposed in~\cite{Kildal}. This approach has the drawback of
strong anisotropy, since the operation of the structure depends
strongly on the incidence angle of the illuminating wave.

In this paper we present a novel idea for realization of an
electromagnetic cloak operating in the microwave region. The
operation of the cloak is based on transmission-line (TL) networks
which can be unloaded or periodically loaded by lumped elements,
such as capacitors and inductors~\cite{Caloz,Eleftheriades}. The
cloak itself is composed of these networks and the wave coming
from the surrounding medium travels through the cloak inside the
transmission-line network (no field mapping is needed). The space
between the neighboring transmission lines of the network is left
undetected, i.e., cloaked from the electromagnetic fields.

The evident drawback of this approach is that a large bulky object
cannot be cloaked due to the periodicity of the transmission-line
network. On the other hand, the cloaked object does not have to be
composed of tiny pieces that are not in contact with each other,
but it can be a three-dimensional mesh of metal or any material.
Also, it is clear that because no field mapping and no exotic
material parameters are required, the cloak design and
manufacturing are fairly straightforward. For simplification we
consider only a two-dimensional cylindrical cloak here (infinite
in the axial direction). However, the same approach can be easily
realized also in three dimensions and for arbitrary shapes, while
preserving full isotropy of (voltage and current) waves traveling
inside the network, see e.g.~\cite{Alitalo}.

\section{Operation of the cloak}

The cloak that we study in this paper is a two-dimensional
simplification of an isotropic three-dimensional cloak, as was
also the design realized in~\cite{Schurig}. Because the cloak is
based on a transmission-line network, it is fairly easy to extend
to three dimensions as well: recently the interest in obtaining
isotropic TL-based metamaterials for superlensing applications has
resulted in the development of different loaded and unloaded
three-dimensional isotropic transmission-line
networks~\cite{Grbic,Hoefer,Alitalo}.

\begin{figure} [b!]
\centering {\epsfig{file=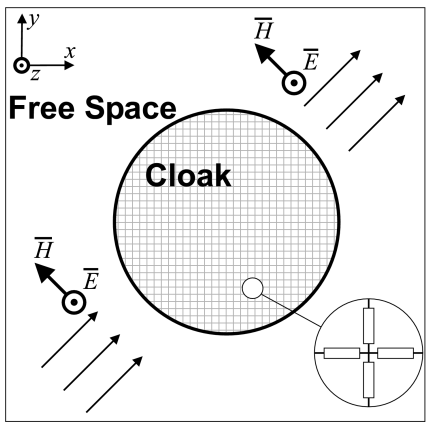, width=0.34\textwidth}}
\caption{A two-dimensional cylindrical electromagnetic cloak. The
cloak operation does not depend on the incidence angle of the
illuminating electromagnetic wave, as long as the network period
is small enough with respect to the wavelength.} \label{TLcloak}
\end{figure}

The cloak design and principle of operation are illustrated in
Fig.~\ref{TLcloak}, where a cylindrically shaped cloak (infinite
in the $z$-direction) is placed in free space. The operation and
optimization of the properties of the cloak naturally depends on
the background medium. In this paper we consider cloaking in free
space (as in~\cite{Leonhardt,Pendry,Alu,Cai,Schurig,Kildal}). The
principle of operation is as following: The incident
electromagnetic wave in free space arrives to the surface of the
cloak, couples into the transmission-line network, and is guided
\textit{through} the cloak that encompasses the object(s) that we
want to be hidden. The cloaked object can be a collection of small
objects, or it can be a mesh of interconnected objects. The only
limitation on the size of the hidden object is that it must fit
inside the network.

There are two preconditions that have to be met in order that the
cloak would operate ideally. Firstly, the wave propagation in the
transmission-line network should mimic the wave propagation in
free space, i.e., the wave propagation should be isotropic and lie
on the same dispersion curve as that in free space. If this
precondition is not satisfied, the cloak will scatter. Secondly,
the transmission-line network should be impedance-matched to free
space in order to prevent reflections from the cloak. The
impedance matching includes also the requirement that there exists
a transition layer around the cloak which couples the waves
propagating in free space into the network. This transition layer
can be realized e.g. with an antenna array.

\section{Dispersion and impedance in loaded and unloaded transmission-line
networks}

To study the propagation inside a transmission-line network, we
derive the dispersion equations for a loaded two-dimensional
transmission-line network, where the loads are defined only as
series impedance $Z$ and shunt admittance $Y$. Here we can use the
same derivation procedure as in~\cite{Alitalo} for a
three-dimensional loaded transmission-line network, where the
loads were defined as $Z=1/(j\omega C)$ and $Y=1/(j\omega L)$, $C$
and $L$ being the capacitance and inductance of the lumped loads.
The dispersion equation for a 3D-case with arbitrary loads $Z$ and
$Y$ reads:

\e \cos(k_x d) + \cos(k_y d) + \cos(k_z d)=
\frac{Y}{2S}-3\frac{K}{S},  \f

and for the 2D-case:

\e \cos(k_x d) + \cos(k_y d) = \frac{Y}{2S}-2\frac{K}{S}, \rm
where  \f

\e S=\frac{Z}{(ZA_{TL}+B_{TL})(ZD_{TL}+B_{TL})-B_{TL}^2},  \f

\e
K=\frac{Z(A_{TL}D_{TL}-B_{TL}C_{TL})(ZA_{TL}+B_{TL})}{[(ZA_{TL}+B_{TL})(ZD_{TL}+B_{TL})-B_{TL}^2]B_{TL}}-\frac{A_{TL}}{B_{TL}},
\f

\e \left[
\begin{array}{ccc}
A_{\rm TL} & B_{\rm TL}  \\
C_{\rm TL} & D_{\rm TL}  \end{array} \right] = \left[
\begin{array}{ccc}
\cos(k_{\rm TL}d/2) & jZ_{\rm TL} \sin(k_{\rm TL}d/2)  \\
jZ_{\rm TL}^{-1} \sin(k_{\rm TL}d/2) & \cos(k_{\rm TL}d/2)
\end{array} \right],  \f $k_i$ is the wavenumber in the network along axis $i$, $d$ is the period and $k_{\rm TL}$ and $Z_{\rm TL}$ are the wavenumber and
impedance of the transmission lines, respectively. The wavenumber
in the network is $k=\sqrt{k_x^2+k_y^2+k_z^2}$.

\begin{figure} [t!]
\centering {\epsfig{file=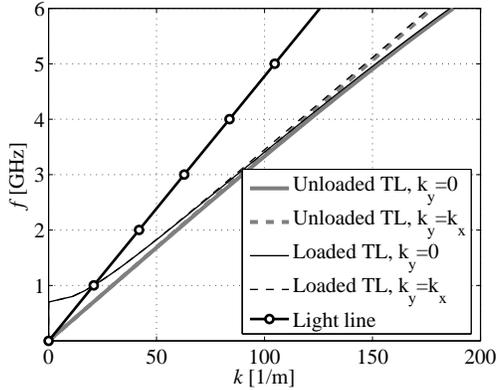,
width=0.45\textwidth}}  \caption{Dispersion in unloaded and loaded
transmission-line networks for axial and diagonal propagation.}
\label{dispersion}
\end{figure}

Obviously the simplest transmission-line network is an unloaded
one. The dispersion equation for an unloaded network is obtained
by using (2) and choosing $Z=0$ and $Y=0$. We have chosen the
dimensions and impedance values for the unloaded network in such a
way that the impedance of the network is approximately equal to
the free space impedance, i.e., 377~$\Omega$, at frequency
$f=1$~GHz. See Table~I for the parameters of the unloaded network
($k_0$ is the free space wavenumber) and Fig.~\ref{dispersion} for
the plotted dispersion curves for axial and diagonal propagation.
From Fig.~\ref{dispersion} it is clear that the phase velocity
$v_{\rm phase}=\omega/k$ in the network slightly differs from that
in free space (light line) for $k>0$.

A capacitively loaded transmission-line network can be designed in
such a way that its phase velocity equals that in free space for a
given frequency point. This can be done by choosing $Z=1/(j\omega
C)$ and $Y=0$ in (2). We have designed this network in such a way
that the propagation in the network and the impedance of the
network are matched with free space at $f=1$~GHz. See Table~I for
the parameters of this capacitively loaded network and
Fig.~\ref{dispersion} for the plotted dispersion curves.

\begin{table}[h!]
\centering \caption{Transmission-line network parameters.}
\begin{tabular}{|c|c|c|c|c|c|}
\hline - & $d$ & $k_{\rm TL}$ & $Z_{\rm TL}$ & $C$ \\  \hline Loaded TL & 8 mm & $k_0$ & 755 $\Omega$ & 2.5 pF\\
 Unloaded TL & 8 mm & $k_0$ & 535 $\Omega$ & -\\\hline
\end{tabular}
\end{table}

\begin{figure} [t!]
\centering {\epsfig{file=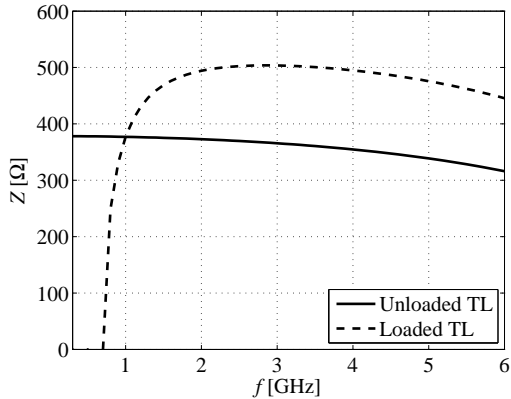,
width=0.45\textwidth}}  \caption{Impedances of the networks,
calculated for axial propagation.} \label{impedance}
\end{figure}

The required isotropy of the networks is achieved if the period is
small compared to the wavelength. From Fig.~\ref{dispersion} we
can conclude that for the presented designs the isotropy is
achieved for both networks approximately below 3~GHz. In
Fig.~\ref{impedance} the impedances of the two studied networks
are calculated using the equations for the network impedances
derived in~\cite{Alitalo}.

From the above discussion, we can conclude that the both types of
networks have certain benefits and drawbacks. Although the phase
velocity can be ideal for the loaded network, the group velocity
is smaller than the speed of light. This means that the loaded
network has the same restriction of cloaking from time-harmonic
fields only, as the designs
in~\cite{Leonhardt,Pendry,Alu,Cai,Schurig}. On the other hand, in
the unloaded network, the phase velocity equals the group
velocity. This enables cloaking from signals. However, the phase
(and group) velocities are not ideal since they do not equal to
those in free space. In the rest of this paper, we will
concentrate on studying the unloaded network, due to the
possibility of signal operation and inherently larger bandwidth.
The larger bandwidth is expected due to the linear dispersion of
the unloaded network. First, we will study the effect of the
aforementioned unideal phase velocity on cloaking and then present
a simple way of matching the unloaded network with free space.

\section{Effect of unideal phase velocity on cloaking}

\subsection{Scattering from the cloak}

To be able to study the effect of unideal phase velocity (which is
related only to the unloaded transmission-line cloak), we consider
a cloak having ideal impedance-matching with the surrounding
medium. The transmission-line network is represented as a
homogeneous material with certain relative permittivity
$\varepsilon_r$ and permeability $\mu_r$ (with their ratio equal
to unity, as in free space). Using (2), it can be found that this
homogeneous material should have relative permittivity and
permeability equal to $\sqrt{2}$. With these parameters the
dispersion curve of our homogeneous material lies exactly on the
dashed curve of Fig.~\ref{dispersion} (i.e., diagonal propagation
in the unloaded network). We can conclude that the homogeneous
material represents the dispersion of the unloaded network very
well for all propagation directions in the region where the
network is isotropic (that is, below 3~GHz). In the following, we
study a cylinder made of a homogeneous material with
$\varepsilon_r=\mu_r=\sqrt{2}$. The cylinder is infinite in the
$z$-direction. The diameter of the cylinder is 128~mm. Note that
here the period is not defined, for the network is modelled with a
homogeneous material.

\begin{figure} [t!]
\centering {\epsfig{file=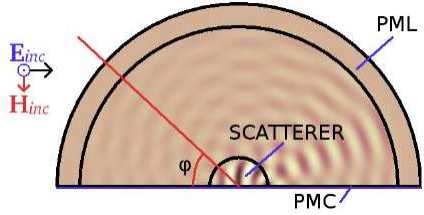, width=0.35\textwidth}}
\caption{Model for calculating the scattering cross section of the
cloak and reference objects.} \label{SCS_model}
\end{figure}

Scattering cross section~\cite{Peterson} is one quantitative
measure for the performance of an idealized cloak. Here the
scattering cross section is calculated numerically using a Finite
Element Method (FEM) based COMSOL Multiphysics software. The
simulation setup is the following: the scatterer is placed in the
center of a cylinder coated with a perfectly matched layer (PML)
and the scatterer (cloak) is illuminated with a plane wave. The
geometrical symmetry allows us to cut the problem in half using a
perfect magnetic conductor (PMC) boundary, see
Fig.~\ref{SCS_model}. From the solved simulation, the scattered
field is calculated in all directions around the scatterer. The
scattered far field $E_s$ is calculated from the near field using
the Stratton-Chu formula on the inner boundary of the
PML~\cite{Lo}. The scattering cross section $\sigma_s$ is obtained
from the scattered far field with

\e \sigma_s(\varphi)=2\pi R\frac{|E_s|^2}{|E_{\rm inc}|^2}, \f
where $\varphi$ is the angle, $R$ is the distance from the center
of the scatterer to the inner boundary of the PML, and $E_{\rm
inc}$ is the incident electric field.

The scattering cross section was calculated for the idealized
cloak and for an array of vertical rods made of a perfect electric
conductor (PEC). The PEC rods have a cross section of
4~mm~$\times$~4~mm (a realizable cloak period has been assumed to
be 8~mm) and the rods are placed in the form of a hexagonally
shaped array. The PEC rod array's outer dimensions are smaller
than the diameter of the simulated homogenized cloak. The PEC rods
constitute the object that could be hidden by the cloak.

\begin{figure} [h!]
\centering {\epsfig{file=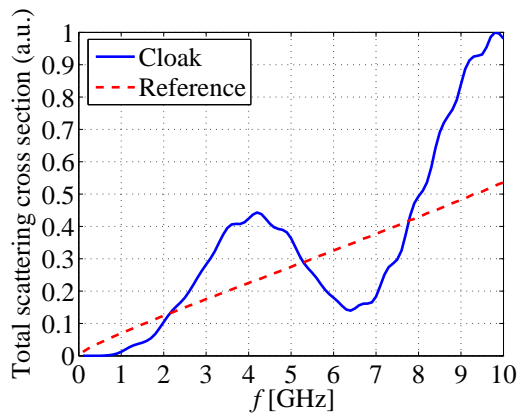,
width=0.45\textwidth}} \caption{Simulated total scattering cross
section from the cloak and the reference object, normalized to the
maximum value of the cloak's total scattering.}
\label{total_scattering}
\end{figure}

The most interesting case to study is the total scattering cross
section, where the scattering cross section is integrated over the
angle $\varphi$. See Fig.~\ref{total_scattering} for the total
scattering of the cloak, calculated from the scattered field
obtained from the simulations. Because the cloak is
impedance-matched to free space, the back scattering from the
(homogeneous) cloak is always below that of the reference object
(this was also verified with simulations). From
Fig.~\ref{total_scattering} we can thus conclude that for some
frequencies the forward scattering from the cloak is very strong,
in some directions even stronger than from the reference object.
From Fig.~\ref{total_scattering} we see that the cloaking
efficiency (total scattering from the cloak as compared to the
reference case) fluctuates quite strongly, whereas the scattering
from the reference object smoothly increases with increasing
frequency. We have found that this phenomenon is due to the phase
matching at the backside of the cloak. For example, at the
frequency 4~GHz, the wave front at the backside of the cloak and
in free space are almost in opposite phase. This leads to strong
scattering from the backside of the cloak.

\begin{figure} [h!]
\centering \subfigure[]{\epsfig{file=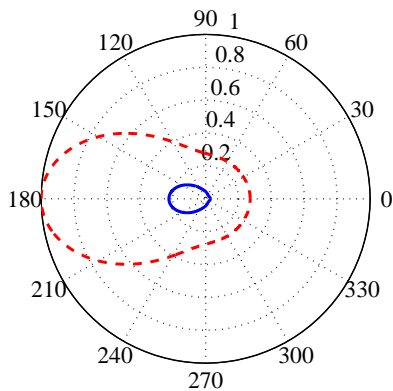,
width=0.5\textwidth}} \subfigure[]{\epsfig{file=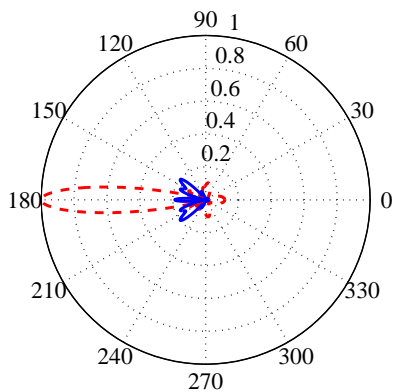,
width=0.5\textwidth}} \caption{Simulated scattered electric field
to different directions from the cloak (solid line) and the
reference object (dashed line) at the frequencies 1~GHz (a) and
6.4~GHz (b). Both plots are normalized to the maximum value of the
reference object scattering. The plane wave that illuminates the
cloak/reference object travels in the direction $\varphi=180^0$}
\label{polar_scattering}
\end{figure}

On the other hand, as the frequency is increased, the phase
difference between the cloak edge and free space becomes smaller
and reaches zero at some frequencies. Therefore the cloaking
efficiency starts to improve after 4~GHz and around 6~GHz there is
a band where the cloak scatters much less than the reference
object. Note that good cloaking efficiency is achieved also at low
frequencies, where the cloak is small compared to the wavelength,
i.e., below 2~GHz in the example case studied here. The scattering
from the cloak and from the reference object to different
directions, at the frequencies 1~GHz and 6.4~GHz, is plotted in
Fig.~\ref{polar_scattering}, from which it is obvious that the
backscattering from the cloak is strongly reduced and also the
forward scattering is considerably lower than that of the
reference object (at those specific frequencies). The reader must
note that in the scattering simulations the cloak is modelled as a
homogeneous material which is fully isotropic at all frequencies.
For the studied transmission-line networks (period $d=8$~mm) the
isotropy is achieved only below the frequency 3~GHz
(approximately), as discussed above. The isotropic region can be
extended to higher frequencies by decreasing the period of the
structure, but this has the drawback of reducing the maximum size
of the cloaked object's inclusions.

\subsection{Cloaking from signals}

To take advantage of the most promising feature of the cloak
proposed in this paper, i.e., cloaking from signals, we have to
use the unloaded transmission-line network. This choice has also
the advantage of larger operation bandwidth, due to the linear
dispersion characteristics and the slowly varying impedance, see
Figs.~\ref{dispersion} and \ref{impedance}. As discussed above,
this choice has the drawback of unideal phase velocity. To obtain
more information about the effect of this unideality on cloaking,
and especially cloaking from signals, we have simulated the cloak
using a Finite-Difference-Time-Domain (FDTD) code. FDTD is
suitable for this task, since it inherently includes the
possibility of signal excitation.

In the FDTD simulations, we have modeled a cloak having the same
diameter as the homogeneous cloak studied above. The diameter of
the cylindrical cloak is therefore $16d=128$~mm and the period of
the interconnecting transmission lines is 8~mm. The cloak has the
same dispersion as shown in Fig.~\ref{dispersion} (unloaded
network). The FDTD code takes into account the anisotropy that
occurs at the higher frequencies. To study only the effect of
unideal dispersion, we have ideal impedance-matching between the
cloak and free space. As a reference object, we again have a mesh
of PEC wires, placed in the position of the cloak. The PEC wires
form a cylindrically shaped mesh that fits inside the studied
cloak. The cloak and the reference object are excited by a signal
with vertically polarized electric field, i.e., $\overline{E}$ is
oriented along the $z$-axis.

\begin{figure} [b!]
\centering \subfigure[]{\epsfig{file=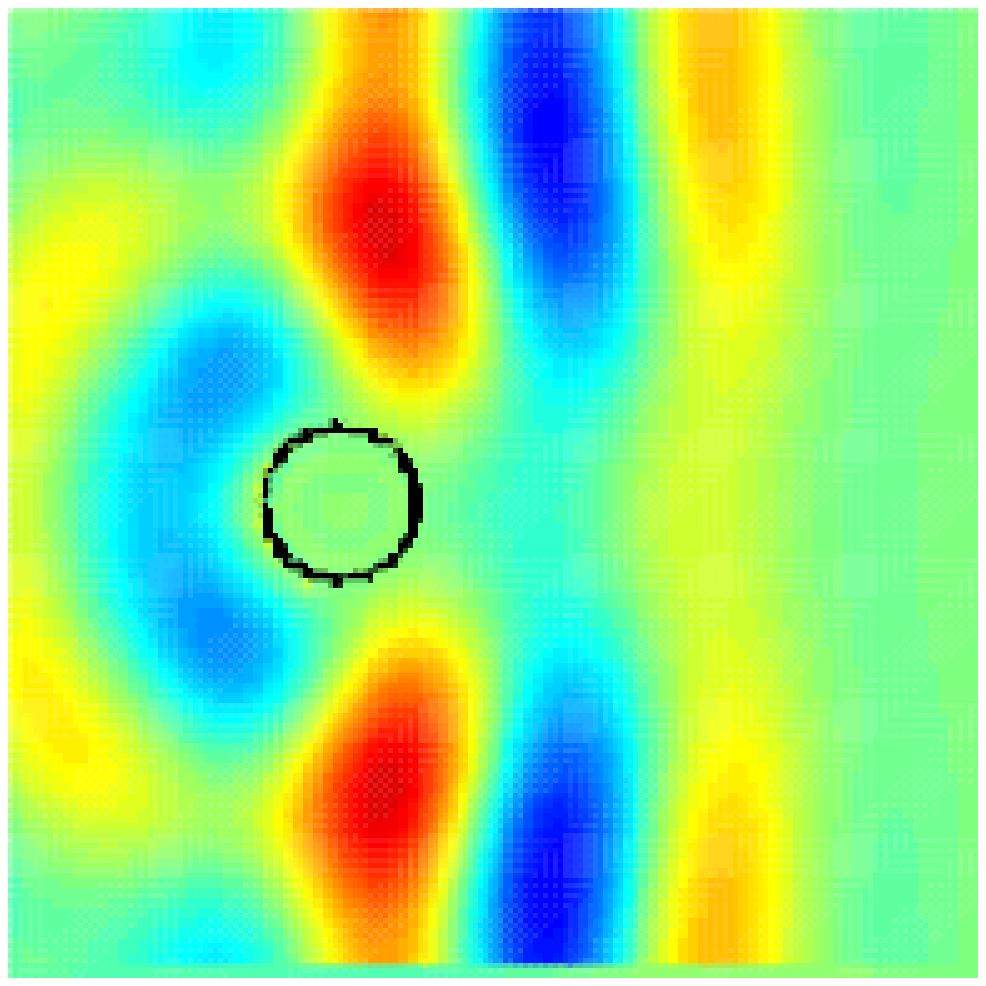,
width=0.225\textwidth}} \subfigure[]{\epsfig{file=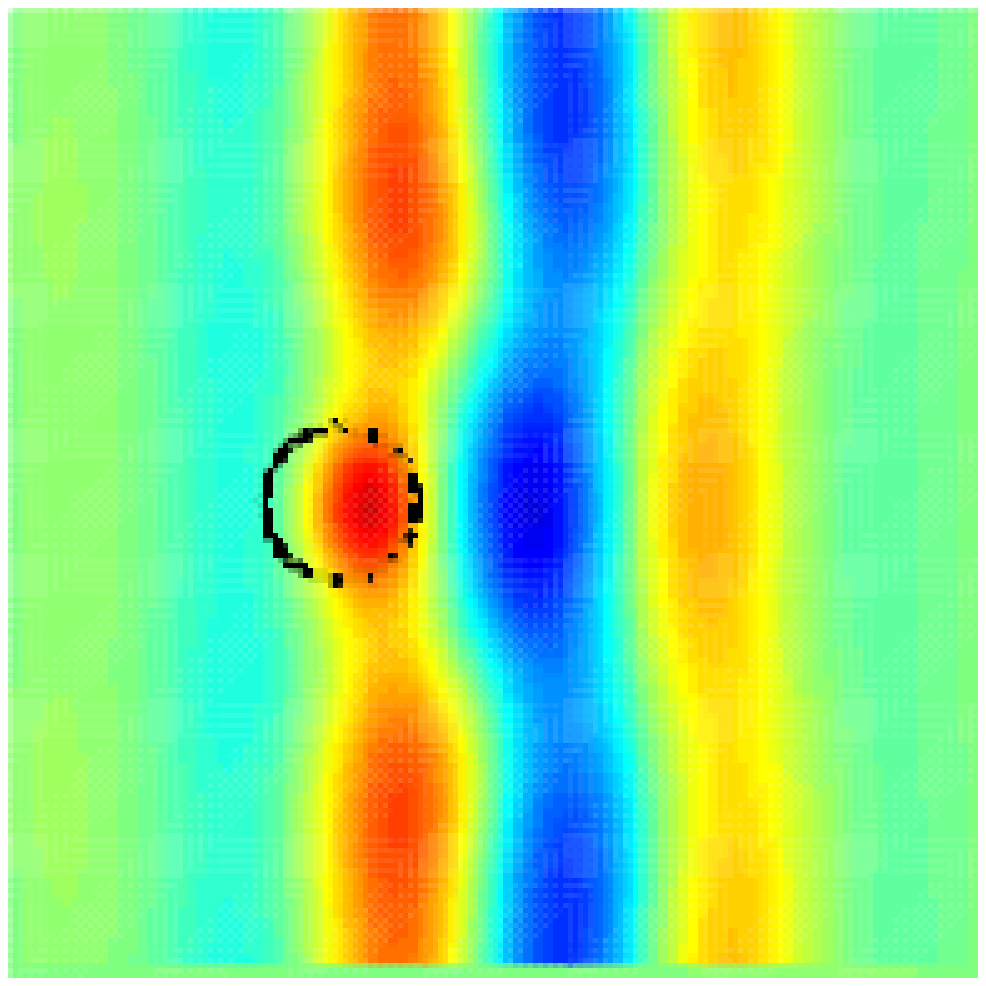,
width=0.225\textwidth}} \subfigure[]{\epsfig{file=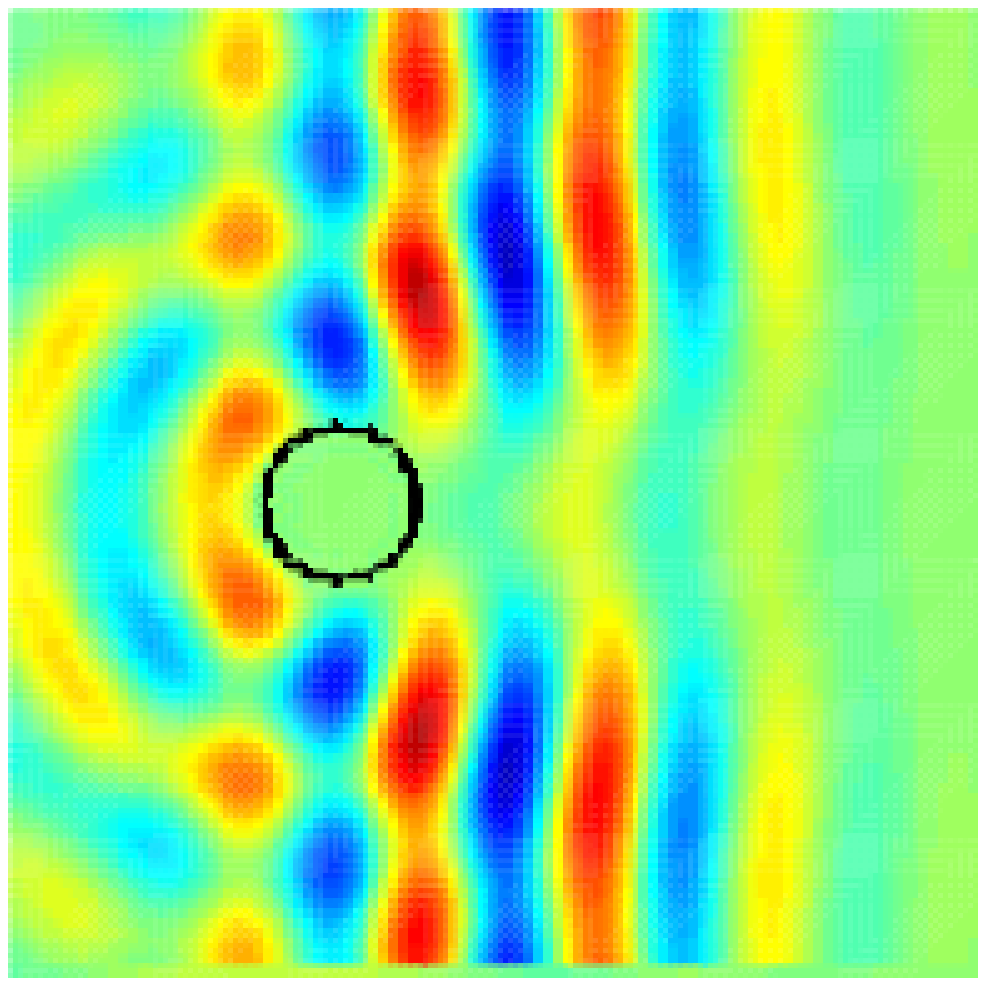,
width=0.225\textwidth}} \subfigure[]{\epsfig{file=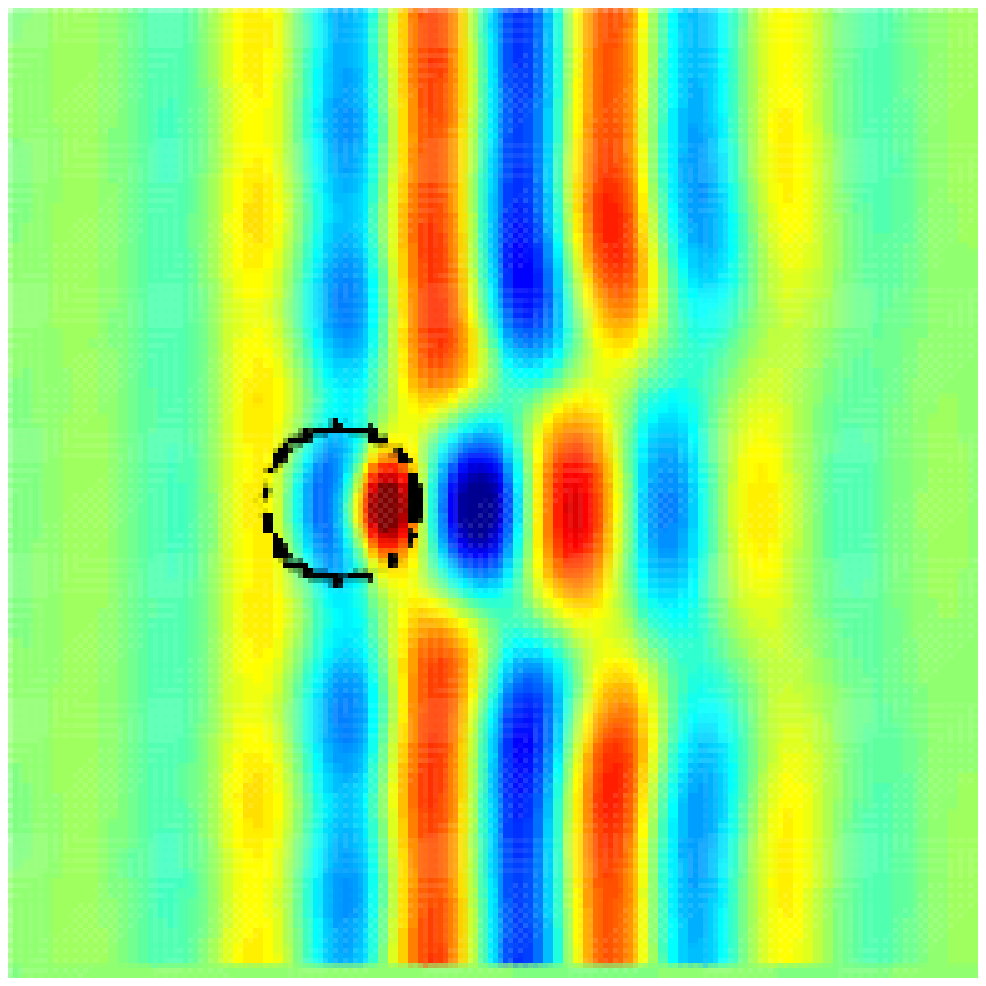,
width=0.225\textwidth}} \subfigure[]{\epsfig{file=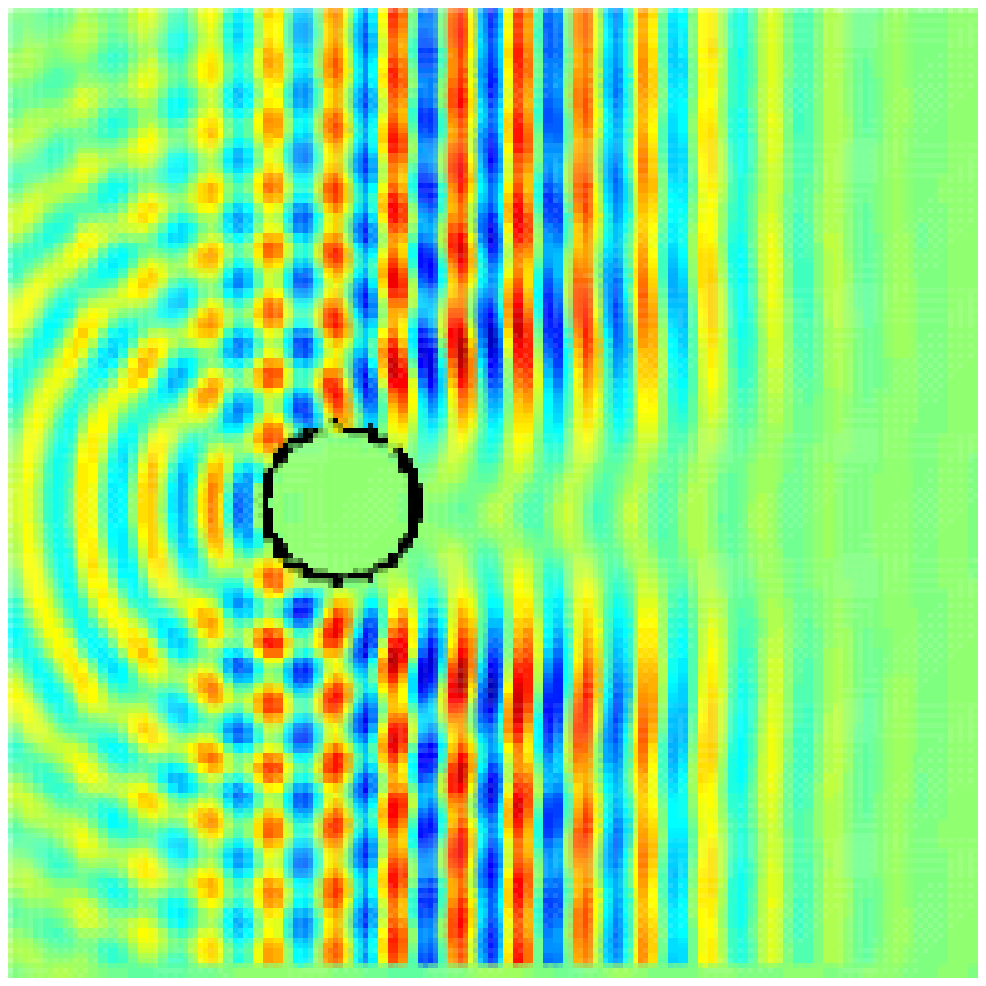,
width=0.225\textwidth}} \subfigure[]{\epsfig{file=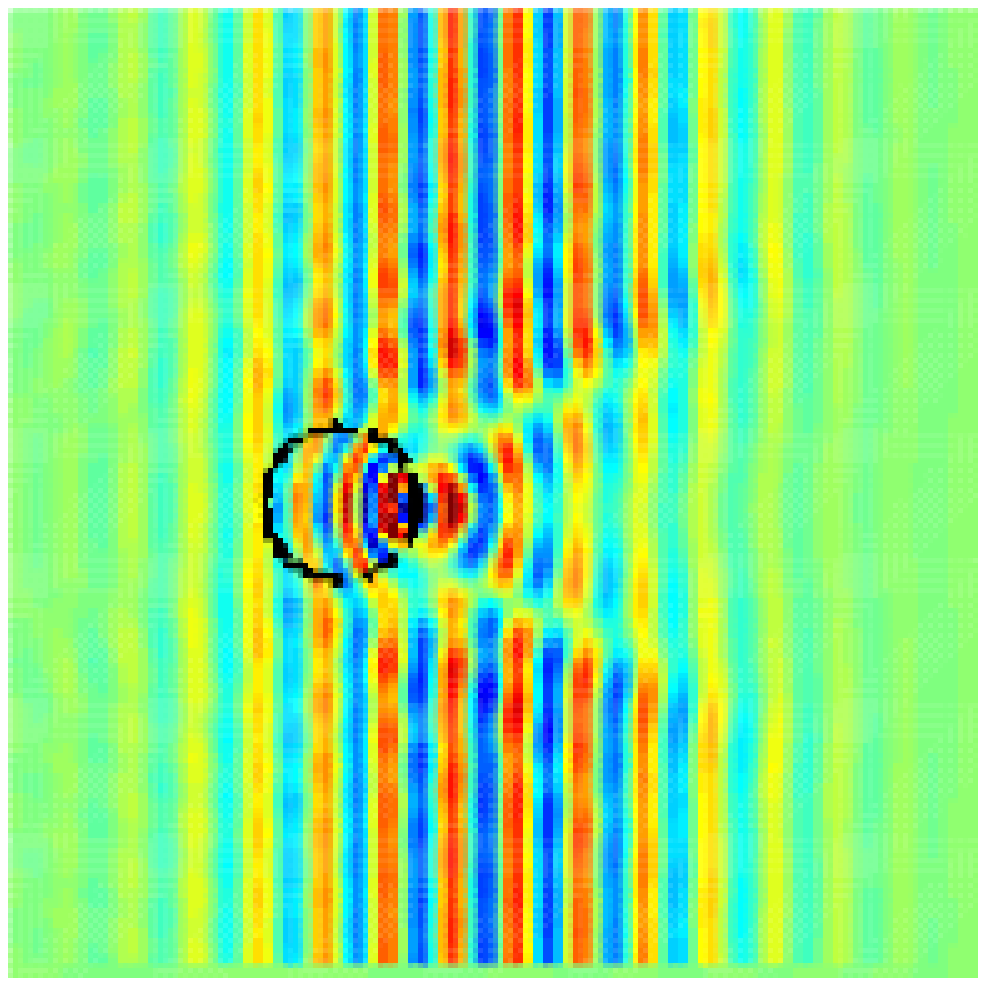,
width=0.225\textwidth}} \caption{Color Online. Electric field
snapshots from FDTD simulations (all normalized to the same
amplitude value). (a) Reference object, $f_c=1$~GHz. (b) Cloak,
$f_c=1$~GHz. (c) Reference object, $f_c=2$~GHz. (d) Cloak,
$f_c=2$~GHz. (e) Reference object, $f_c=5.5$~GHz. (f) Cloak,
$f_c=5.5$~GHz.} \label{FDTD}
\end{figure}

According to Fig.~\ref{total_scattering}, we can expect good
cloaking (as compared to the reference object) at frequencies
below 2~GHz, and e.g., in the band from 5.5~GHz to 7.5~GHz. See
Fig.~\ref{FDTD} for the simulation results for pulses with center
frequencies at 1~GHz, 2~GHz, and 5.5~GHz. In Fig.~\ref{FDTD}
snapshots of the electric field distributions are taken at a time
step when the signal has just passed the cloak/reference object,
i.e., the signal comes from the left and moves to the right in
Fig.~\ref{FDTD}. The signal bandwidths used in the FDTD
simulations are relatively large: in Figs.~\ref{FDTD}(a),(b) the
relative half-power pulse width is 52~\% with center frequency
$f_c=1$~GHz, in Figs.~\ref{FDTD}(c),(d) it is 26~\% with center
frequency $f_c=2$~GHz and in Figs.~\ref{FDTD}(e),(f) it is 9.5~\%
with center frequency $f_c=5.5$~GHz.

From Fig.~\ref{FDTD} we see that at all the shown frequencies, the
backscattering from the cloak is zero, and that inside the cloak,
the waves are ``delayed,'' with respect to the waves propagating
in free space (this is due to the shorter wavelength inside the
cloak). Because of this, some forward scattering from the cloak
occurs, although the shadow generated by the cloak is always well
mitigated, as compared to the reference object. The FDTD
simulations seem to confirm the cloaking effect, studied for the
time-harmonic case in Fig.~\ref{total_scattering}, also for signal
excitation. The pulse with center frequency of 5.5~GHz, presented
in Figs.~\ref{FDTD}(e),(f) is already at the frequency range where
the network is anisotropic. The transmission-line network in FDTD
simulations is not capable of supporting higher frequencies,
because the wavelength becomes too short compared to the period of
the network. Despite of the anisotropy, Fig.~\ref{FDTD}(f)
demonstrates the benefit of the phase matching inside and outside
the cloak: because the waves that emerge from the backside of the
cloak are in phase with the waves outside the cloak, the
wavefronts at the back of the cloak prevent the formation of a
strong shadow.

It is clear that the best cloaking effect (cloak scattering versus
the reference object scattering) is always achieved at low
frequencies. The phase matching technique can be used if the cloak
must be large compared to the wavelength.

\section{Simulation of a realizable cloak}

In this section we present a simple way of matching the
transmission-line network to free space. Furthermore, we verify
the operation of the cloak by numerical full-wave simulations
using Ansoft's High Frequency Structure Simulator (HFSS). The
previous sections of this paper were devoted to study of the
unideal propagation characteristics of the unloaded network while
assuming perfect impedance matching between the cloak and the
surrounding medium. In this section we study a transversally
infinite slab made of two-dimensional unloaded transmission-line
networks, thus effectively studying only the transition from free
space to the cloak and vice versa. The reader should notice here
that unlike in the previous section, the thickness of the slab
does not have any effect on the phase matching between the waves
travelling inside and outside of the cloak for the cloak is
infinite in the transversal direction. However, since the
wavelength inside the slab differs from that in free space we
expect to see Fabry-Perot type resonances in the reflection and
transmission coefficients due to the finite electrical length of
the slab (along the $x$-axis).

\begin{figure} [b!]
\centering \epsfig{file=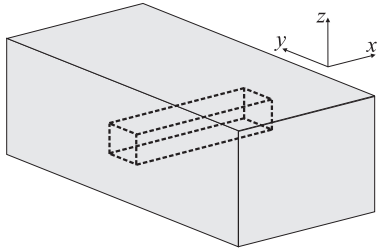, width=0.3\textwidth}
\caption{The modeled slab made of two-dimensional
transmission-line networks which are placed on top of each other.
The slab is infinite in the $y$- and $z$-directions. The dashed
line shows a single ``unit cell'' of the slab, which can be easily
modelled with full-wave simulators using periodic boundaries.}
\label{modelslab}
\end{figure}

The periodical boundary conditions allow us to simplify the
simulation model: We need to model only one ``unit cell'' of the
infinite slab as shown in Fig.~\ref{modelslab}. The slab shown in
Fig.~\ref{modelslab} is infinite in the $y$- and $z$-directions,
but along the $x$-axis it is finite with thickness of
$16d=128$~mm.

Based on the simulated scattering and the FDTD results, we have
decided to tune the network impedance in order to obtain the
impedance matching at higher frequencies (around 6~GHz).
Effectively this means that the impedance of single sections of
transmission lines should be increased to raise the curve shown in
Fig.~\ref{impedance} to obtain the impedance of 377~$\Omega$ at a
frequency near 6~GHz. Using the impedance equations
from~\cite{Alitalo} we have found that the transmission-line
impedance of 600~$\Omega$ results in the network impedance of
377~$\Omega$ at approximately 5.2~GHz. As the transmission lines,
we have chosen to use parallel metal strips due to the large range
of impedance values available and the simple geometry and easy
manufacturing of this type of transmission line. Using the simple
parallel-plate approximation, we have calculated the width $w$ and
height $h$ of the transmission line to be 1.257~mm and 2~mm,
respectively. The metal strips are modelled as infinitely thin and
perfectly conducting to reduce the simulation time.

\begin{figure} [t!]
\centering \subfigure[]{\epsfig{file=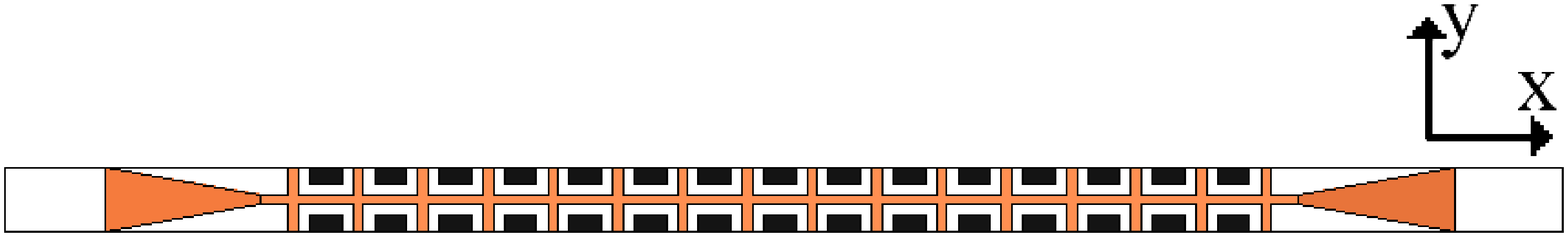,
width=0.475\textwidth}}
 \subfigure[]{\epsfig{file=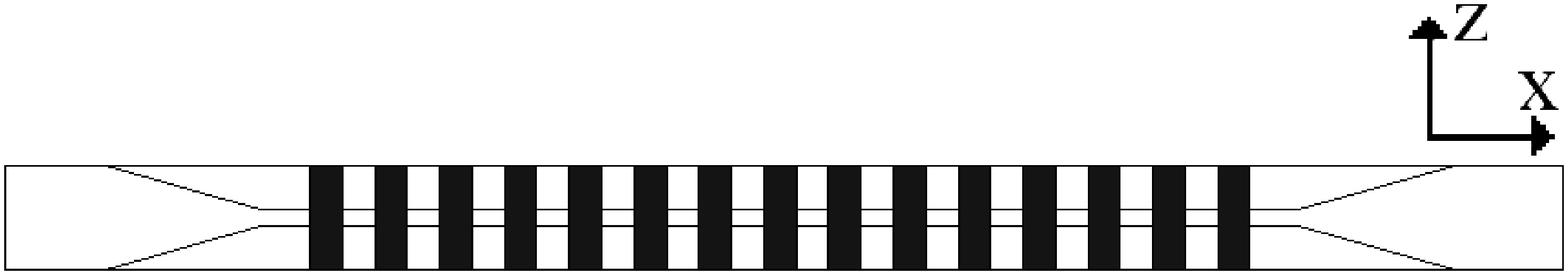,
width=0.475\textwidth}} \caption{HFSS model of the cloak with
reference object inside (a) viewed from the top and (b) viewed
from the side. The reference object is illustrated with black
color.} \label{model}
\end{figure}

Matched transition from free space to the cloak and vice versa can
be obtained by extending the transmission lines at the both sides
of the cloak, while preserving the impedance of the transmission
lines. This way the both slab surfaces are covered with small
sections of transmission lines, effectively operating as antennas
with their input impedance equal to the free space impedance. See
Fig.~\ref{model} for the HFSS simulation model of the ``unit
cell'' of the slab.

As shown in Fig.~\ref{model}, the antennas cover almost all of the
area of the unit cell in the $yz$-plane, but in fact, there is a
tiny gap left between the antennas of the neighboring unit cells
(approximately 2~$\mu m$ in the $z$-direction and 120~$\mu m$ in
the $y$-direction). The width and height of the simulated unit
cell are 8~mm and 12.73~mm, respectively, to preserve the correct
impedance of the transmission lines ($12.73/8\approx 2/1.257$). We
have also placed a reference object that we want to cloak, inside
the network, see Fig.~\ref{model}. This reference object is an
array of rods made of PEC, as in the scattering simulations. The
rods have a cross-section of 4~mm~$\times$~4~mm and they are
effectively infinite along the $z$-axis.

\begin{figure} [t!]
\centering \subfigure[]{\epsfig{file=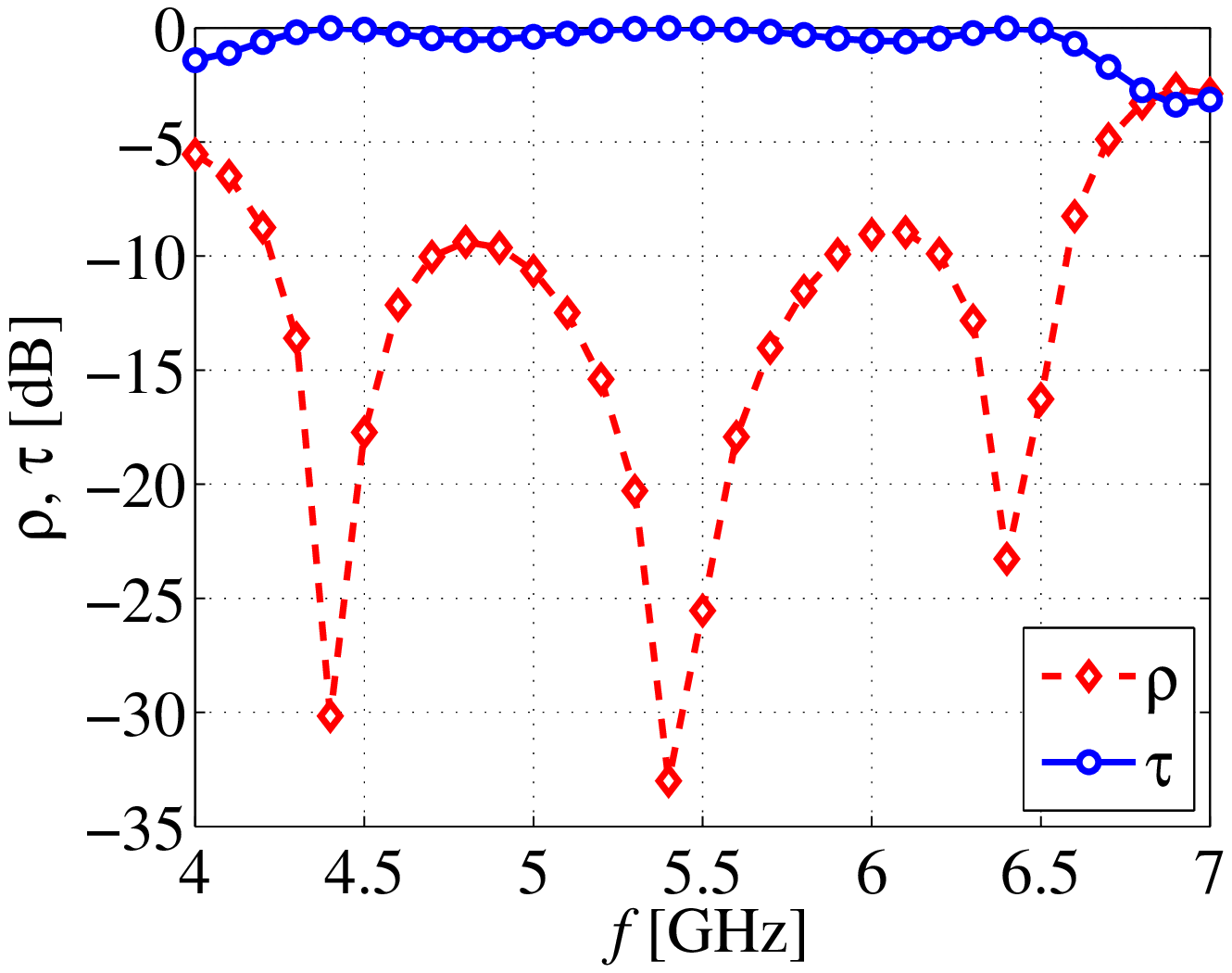,
width=0.4\textwidth}}
 \subfigure[]{\epsfig{file=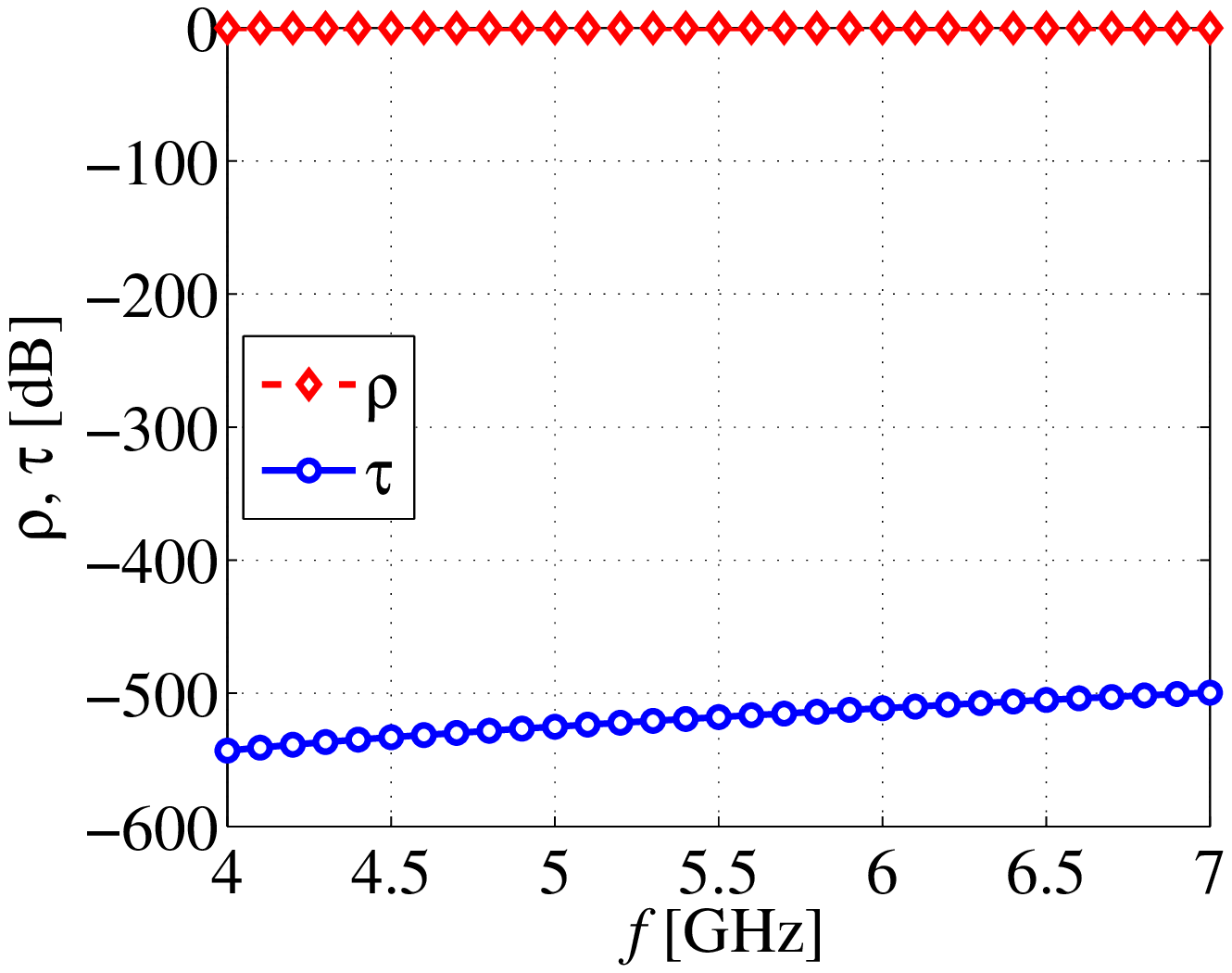,
width=0.4\textwidth}} \caption{Simulated reflection and
transmission for (a) cloak slab with reference object inside and
(b) reference object. A plane wave illuminates the cloak slab with
normal incidence angle.} \label{reflection_transmission}
\end{figure}

The cloak is excited with a plane wave having the electric field
parallel to the $z$-axis. First, we have studied the case of the
normal incidence angle (magnetic field $\overline{H}$ is parallel
to the $y$-axis). See Fig.~\ref{reflection_transmission}(a) for
the simulated reflection ($\rho$) and transmission ($\tau$)
coefficients for the cloak with the reference object inside and
Fig.~\ref{reflection_transmission}(b) for $\rho$ and $\tau$ in the
case of the reference object only.

From Fig.~\ref{reflection_transmission} it is obvious that the
reference object behaves effectively as a solid metal slab
reflecting basically all of the incident field, whereas the same
object placed inside the cloak is almost transparent to the
incident field. We believe that the resonances seen in the
reflection and transmission through the cloak are due to the
finite thickness of the cloak, as discussed above. The minimum of
the envelope of the reflection coefficient in
Fig.~\ref{reflection_transmission}(a) occurs at approximately
5.4~GHz, which is close to the expected value of 5.2~GHz, which
was obtained from the analytical expressions for the network
impedance.

\begin{figure} [t!]
\centering {\epsfig{file=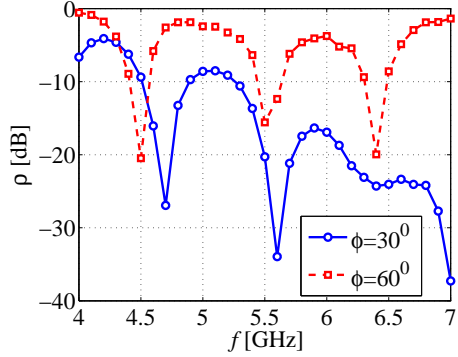, width=0.4\textwidth}}
\caption{Simulated reflection coefficients for the cloak with
reference object inside. Oblique incidence angles of $\phi=30^0$
and $\phi=60^0$ are considered.} \label{reflection_obl}
\end{figure}

The effect of oblique incidence angle was also considered for two
cases: $\phi=30^0$ and $\phi=60^0$, where $\phi$ is the angle in
the $xy$-plane, i.e., the electric field $\overline{E}$ is always
directed along the $z$-axis. See Fig.~\ref{reflection_obl} for the
simulation results of the reflection coefficient $\rho$ as a
function of the frequency for these two incidence angles.

Note that the positions of the resonance frequencies seen in the
reflection coefficient are affected by the oblique incidence due
to the shift in the Fabry-Perot resonances: the effective
thickness of the slab changes as the incidence angle changes. At
frequencies above 3~GHz the transmission-line networks studied in
this paper are anisotropic, which also changes the electrical
thickness of the slab for different incidence angles.
Nevertheless, for the frequencies 5.5~GHz and 6.4~GHz, the
reflection coefficient remains very low for all the studied
incidence angles.

\section{Conclusions}

We have shown that transmission lines can be used to cloak objects
from electromagnetic fields. The object or objects that can be
cloaked have to be smaller than the period of the cloak itself,
which is composed of either loaded or unloaded transmission-line
networks. The propagation characteristics of the transmission-line
network have to mimic the propagation in the medium that surrounds
the cloak. If this medium is free space, the optimal propagation
characteristics of the cloak cannot be achieved with an unloaded
network. A periodically loaded transmission-line network on the
other hand can be designed to have ideal propagation
characteristics, but the operation of this cloak is very
narrowband and it suffers from similar unidealities as the cloak
designs presented in the literature. The unloaded
transmission-line networks seem to be more prospective for
applications that require simple manufacturing and large
bandwidth.

We have thoroughly studied the effect of unideal propagation
characteristics on the cloaking effect and have shown that when
the cloak is small enough as compared to the wavelength, good
cloaking efficiencies are achieved. Also, at higher frequencies
the cloaking effect is achieved at frequencies that depend on the
physical size of the cloak. This is due to the phase matching of
the wave at the backside of the cloak and the wave in the
surrounding medium. We have shown how a realizable
transmission-line network can be matched to free space and have
presented full-wave simulations for a slab made of such a network
placed in free space.

\section*{Acknowledgements}

This work was initially started as a student project work at the
TKK Radio Laboratory and the TKK Electromagnetics Laboratory
within a postgraduate course ``Metamaterials in Electromagnetics
and Radio Engineering.'' The authors wish to thank
Mr.~A.~Karttunen, Mr.~G.~Molera, Mr.~H.~Rimminen,
Mr.~M.~Vaaja and Prof.~A.~Sihvola for useful discussions during that course.\\
This work has been partially funded by the Academy of Finland and
TEKES through the Center-of-Excellence program. Pekka Alitalo
wishes to thank the Graduate School in Electronics,
Telecommunications and Automation (GETA) and the Nokia Foundation
for financial support. Liisi Jylh\"{a} would like to thank GETA
for financial support.

\end{document}